# Room-temperature ferromagnetic insulating state in highly cation-ordered epitaxial oxide double perovskite

Changhee Sohn, Elizabeth Skoropata, Yongseong Choi, Xiang Gao, Ankur Rastogi, Amanda Huon, Michael A. McGuire, Lauren Nuckols, Yanwen Zhang, John W. Freeland, Daniel Haskel, and Ho Nyung Lee*

Dr. C. Sohn, Dr. E. Skoropata, Dr. X. Gao, Dr. A. Rastogi, A. Huon, Dr. M. A. McGuire, Dr. Y. Zhang, Dr. H. N. Lee

Materials Science and Technology Division, Oak Ridge National Laboratory, Oak Ridge, Tennessee 37831, USA

E-mail: hnlee@ornl.gov

Dr. Y. Choi, Dr. J. W. Freeland, Dr. D. Haskel

Advanced Photon Source, Argonne National Laboratory, Argonne IL 60439, USA

A. Huon

Drexel University, Department of Materials Science & Engineering, Philadelphia, Pennsylvania 19104, USA

L. Nuckols, Dr. Y. Zhang

Department of Materials Science and Engineering, University of Tennessee, Knoxville, Tennessee 37996, USA

*This manuscript has been authored by UT-Battelle, LLC under Contract No. DE-AC05-00OR22725 with the U.S. Department of Energy. The United States Government retains and the publisher, by accepting the article for publication, acknowledges that the United States Government retains a non-exclusive, paid-up, irrevocable, world-wide license to publish or reproduce the published form of this manuscript, or allow others to do so, for United States Government purposes. The Department of Energy will provide public access to these results of federally sponsored research in accordance with the DOE Public Access Plan (http://energy.gov/downloads/doe-public-access-plan).*

**Abstract**

Ferromagnetic insulators (FMIs) are one of the most important components in developing dissipationless electronic and spintronic devices. However, since ferromagnetism generally accompanies metallicity, FMIs are innately rare to find in nature. Here, novel room-temperature FMI films are epitaxially synthesized by deliberate control of the ratio of two B-site cations in the double perovskite $Sr_2FeReO_6$. In contrast to the known ferromagnetic metallic phase in stoichiometric $Sr_2FeReO_6$, a FMI state with a high Curie temperature ($T_c \approx 400$ K) and a large saturation magnetization ($M_S \approx 1.8$ μB/f.u.) is found in highly cation-ordered Fe-rich phases. The stabilization of the FMI state is attributed to the formation of extra $Fe^{3+}$–$Fe^{3+}$ and $Fe^{3+}$–$Re^{6+}$ bonding states, which originate from the excess Fe. The emerging FMI state by controlling cations in the epitaxial oxide perovskites opens the door to developing novel oxide quantum materials & heterostructures.

High Curie temperature ($T_c$) ferromagnetic insulators (FMIs) are critical for realizing dissipationless quantum electronic/spintronic devices and solid-state quantum computing.[1] FMIs are essential to filter electronic charges to generate pure spin currents, to manipulate spins in nonmagnetic layers by magnetic proximity effect,[2-4] and to create a quantum anomalous Hall effect in combination with topological insulators.[5, 6] Unfortunately, the ferromagnetism generally accompanies metallicity in nature. The most common mechanism for the ferromagnetism is the RKKY interaction,[7-9] where itinerant electrons mediate FM interactions among local spins. Such interaction, therefore, is completely suppressed in insulating materials, and an antiferromagnetic ground state generally emerges. While oxide perovskites are promising materials for future technologies owing to the rich correlated phenomena, such as high $T_C$ superconductivity, colossal magnetoresistance, ferroelectricity, and metal-insulator transition, few perovskite FMIs are known and their magnetic transitions occur at low temperature ($T$).[10]

The 3$d$–5$d$ double perovskite system (A$_2$B'B"O$_6$) is a promising candidate for FMIs because of the unique mechanism of their ferromagnetism and different nature of 3$d$ and 5$d$ orbitals. Double perovskites are compatible with other oxide perovskites owing to the similar crystallographic structure. The two different B-site cations are known to order in the checkerboard type as shown in **Figure. 1**a. Since some of these compounds are ferromagnetic metals (FMMs) above room $T$, they have been considered as a promising candidate for spintronic devices.[11-16] The uniqueness of ferromagnetism in these materials is that it originates from the so-called hybridization mechanism instead of the RKKY interaction.[17] In the hybridization mechanism, high-$T$ ferromagnetism is stabilized by large exchange splitting of one of the two magnetic ions and spin-dependent hybridization between the two B-site $d$ orbitals.[11, 18] Since itinerant electrons do not play a role in inducing ferromagnetism, the latter can be robust against the depletion of free carriers. Among those double perovskites, particular attention is paid to systems composed of 3$d$ and 5$d$ transition metal ions, which have potential to show a wide range of electronic and magnetic ground states owing to the cooperative/competitive interactions between strongly localized 3$d$ orbitals and extended/spin-orbit coupled 5$d$ orbitals.

Here, we report a way to synthesize highly insulating $Sr_2FeReO_6$ double perovskite epitaxial films with high $T_c$ ferromagnetism ($\approx$ 400 K) by properly controlling the degree of cation ordering and the ratio of Fe and Re. In addition to the original FMM state, which can be found when the Fe/Re ratio is balanced, FMI states are obtained in the growth regime of Fe-rich and cation-ordered phases. The room-$T$ sheet resistance of insulating films are about three orders of magnitude larger than that of metallic films. X-ray absorption and optical spectroscopy further revealed that the formation of $Fe^{3+}$–$Fe^{3+}$ and $Fe^{3+}$–$Re^{6+}$ bonding is responsible for the insulating electronic structure in Fe-rich $Sr_2FeReO_6$ films.

We used pulsed laser deposition to synthesize high-quality $Sr_2FeReO_6$ epitaxial thin films on (001) $SrTiO_3$ substrates, which are thermally and chemically treated to achieve atomically flat, $TiO_2$-terminated surfaces. For crystallographic parameters, we treated $Sr_2FeReO_6$ as pseudocubic with a slight tetragonal distortion ($a_{pc} = b_{pc}$ = 7.864 Å and $c_{pc}$ = 7.901 Å) to match the crystallographic direction with that of the cubic substrate.[19] It is known that the growth of double perovskite thin films is sensitive to many growth parameters, including oxygen pressure ($P_{O2}$), substrate $T$, and growth rate.[20-24] Among many growth parameters, we found that $P_{O2}$ was the most critical growth parameter affecting the cation ordering and ratio in $Sr_2FeReO_6$ films. Figure 1b exhibits x-ray diffraction (XRD) $\theta$–2$\theta$ scans of $Sr_2FeReO_6$ films grown under various $P_{O2}$ at fixed $T$ = 775 °C. Note that, within the whole range of $P_{O2}$, high-quality epitaxial films were obtained. However, as shown in Figure 1c, the evidence of the cation ordering and its strong $P_{O2}$ dependence can be found by the appearance of the Sr$_2$FeReO$_6$ 111 peak. When B-site cations are ordered in the checkerboard type, they are alternatively stacked along the [111] direction as highlighted by the red and blue planes in Figure 1a. This ordering is confirmed by XRD off-axis scans at $\psi = 54.7°$ near the Sr$_2$FeReO$_6$ 111 peak. As shown in Figure 1c, the 111 peak is only observable for films grown with a $P_{O2}$ higher than 15 mTorr. Note that not all the epitaxial films revealed the cation ordering based on the off-axis XRD scans.

The cation ordering is further confirmed by scanning transmission electron microscopy (STEM). The schematic drawing in Figure 1d shows a projected structure of Sr$_2$FeReO$_6$ along the [1-10] direction. The two B-site cations are ordered along the projected direction, so that they can be resolved by high-angle annular dark field (HAADF) imaging.[25] In HAADF images, Re ions (atomic number Z = 75) give the brightest intensity, while the intensity of Sr (Z = 38) and Fe (Z = 26) ions is less intense as they are lighter than Re. The middle panel of Figure 1d is a HAADF image of a cation-ordered film grown at 20 mTorr. The clear ordering highlighted by a blue diamond is observed as expected from the Re position of ordered double perovskite shown in the left-side schematic drawing. In the HAADF image of the disordered film grown at 1 mTorr (right panel of Figure 1d), on the other hand, such ordering was destroyed, which can be confirmed by a uniform contrast for all the B sites. This observation is consistent with the disappearance of the 111-ordering peak in XRD scans for the samples grown in low $P_{O2}$ (< 15 mTorr).

In addition to B-site cation ordering, we performed Rutherford back-scattering spectrometry (RBS) experiments to measure the Re/Fe ratio as a function of $P_{O2}$ as shown in Figure 1e. A clear linear trend guided by the blue solid line was observed; the Re/Fe ratio decreases as the growth $P_{O2}$ increases. The ideal 1:1 Re/Fe ratio was achieved only in the very narrow $P_{O2}$ range from 10 to 20 mTorr. By considering the cation ordering shown in Figure 1c, we conclude that the oxygen pressure must be well balanced in three regimes, i.e., low $P_{O2}$ (<15 mTorr), $P_{O2}$ near 20 mTorr (15 ~ 20 mTorr), and high $P_{O2}$ (≥ 25 mTorr), to make disordered/Re-rich, ordered/stoichiometric, and ordered/Fe-rich phases, respectively.

We investigated the influence of different Re/Fe ratio and ordering on the magnetic ground state of the $Sr_2FeReO_6$ films grown under different $P_{O2}$ conditions. **Figures 2**a and b display magnetization (*M*) versus *T* curves and *M* versus magnetic field (*H*) curves of our $Sr_2FeReO_6$ films, respectively. The cooling and measuring field for *M*(*T*) curves was 1000 Oe, and the *M*(*H*) curves were measured at 10 K. We first focused on the ordered/stoichiometric film (indicated by the green line). The *M*(*T*) curve was consistent with typical of long-range ferromagnetic ordering. To check the $T_c$ of the films, we further obtained saturated magnetization ($M_S$) data at high *T*. The results are shown as open circles in the Figure 2a. From the data, a $T_c$ near 400 K was determined to be similar to previous reports of 390-410 K for bulk samples.[26-28] The *M*(*H*) curve in Figure 2b for the ordered/stoichiometric film shows a clear hysteresis behavior as well. The $M_S$ of 2.1 ± 0.1 $\mu_B$/f.u. lower than the ideal value 3.33 $\mu_B$/f.u. was likely due to potential antisite disorder in films.[27] Note that the magnetism in $Sr_2FeReO_6$ is known to be more sensitive to the amount of antisite disorders than other double perovskites.[25]

This high-*T* ferromagnetic ground state was very sensitive to the cation ordering but was robust against the cation ratio change. For all of the films with cation ordering and non-stoichiometry, e.g., the film grown at 100 mTorr, we observed a clear magnetic transition near 400 K as well as hysteresis behavior with the $M_S$ approximately 1.8 $\mu_B$/f.u.. For cation-disordered films (e.g., films grown in 1 and 10 mTorr $P_{O2}$), on the other hand, a long-range magnetic ordering is completely destroyed, as evidenced in no *T*-dependence in *M*(*T*) curves and lack of hysteresis in $M(H)$. A small but finite value of magnetization in disordered samples at high fields may imply the presence of the short-range magnetic ordering or spin-glass behaviors.

Unlike the magnetization, the transport properties were more sensitive to the cation ratio, leading to a metal-insulator transition while preserving the ferromagnetic ground state. Figure 2c shows the $T$-dependent sheet resistance, $R_S(T)$, of the $Sr_2FeReO_6$ films measured from 10 to 400 K. A clear metal-insulator transition was observed as the growth $P_{O2}$ was increased. The electronic ground state was metallic for disordered/Re-rich ($P_{O2}$ = 1, 10 mTorr) and ordered/stoichiometric ($P_{O2}$ = 20 mTorr) samples because their $R_S(T)$ showed a negative or negligible $T$ dependence with decreasing $T$. Small upturns in $R_S(T)$ at low $T$ come from the impurity scattering. For ordered/Fe-rich samples ($P_{O2} \geq 25$ mTorr), on the other hand, a clear insulating behavior was observed. This monotonic increase of resistivity as a function of the growth $P_{O2}$ indicates that the transport properties of $Sr_2FeReO_6$ are highly sensitive to the Re/Fe ratio. Figure 2d shows $R_S$(300 K) as a function of Re/Fe ratio determined from the RBS technique. There is a dramatic increase of resistivity (indicated by the blue line) when Re/Fe ratio decreases below 1, indicating that the transport behaviors are very sensitive to Re ion deficiency. It should be noted that variation of room-$T$ resistivity is about three orders of magnitude. The observed metal-insulator transition in our single crystalline films is more dramatic than that reported for polycrystalline $Sr_2Fe_{1+x}Re_{1-x}O_6$,[29] which can be attributed to the absence of metallic grain boundaries in our film that often found in polycrystalline films. Note that the metal-insulator transition with robust ferromagnetism induced by excessive Fe provides a basis for creating a novel high-$T$ FMI in contrast to the metallic ground state of bulk materials. Figure 2e summarizes the electronic and magnetic phases found in our $Sr_2FeReO_6$ epitaxial films as a function of $P_{O2}$ during growth, which includes paramagnetic metal (PMM), FMM, and FMI states.

We used x-ray absorption spectroscopy (XAS) to gain a deeper understanding of the FMI state in Fe-rich $Sr_2FeReO_6$ films. **Figure 3**a shows total electron yield (TEY) XAS spectra near the Fe $L_3$ and $L_2$ edges of a Re-rich PMM, stoichiometric FMM, and Fe-rich FMI. The spectral features from metallic and insulating films were overall similar to those of octahedral $Fe^{3+}$. The two-peak structure in Fe $L_3$ and $L_2$ edges originates from the transitions from $2p$ core levels to empty $3d$ $t_{2g}$ and $e_g$ orbitals. The only noticeable change we found in the FMI film is that the peaks for unoccupied $t_{2g}$ orbitals are better resolved and have a higher intensity than those from metallic films. Similar trends have been observed in fluorescence yield (FY) spectra (the inset of Figure 3a), indicating that such a spectral change did not originate from the surfaces. While the formation of $Fe^{2+}$ ions may explain the increased intensity at the $t_{2g}$ peaks, such formation is unlikely in Fe-rich phases, where Fe should prefer to be $Fe^{4+}$ for charge neutrality. The formation of $Fe^{4+}$ ions, known to broaden $L$-edge XAS spectra significantly, is also inconsistent with our observation that the spectrum of the Fe-rich FMI film shows shaper $t_{2g}$ and $e_g$ peaks than those of the PMM and FMM films. In addition, a previous Mössbauer study on $Sr_2Fe_{1+x}Re_{1-x}O_6$ polycrystalline samples revealed the robust trivalent $Fe^{3+}$ state is maintained in Fe-rich samples.[29] We found that the decrease in the hybridization between the Fe $3d$ and Re $5d$ orbitals owing to the Re deficiency can explain this spectral change. In this double perovskite, there is strong hybridization between Fe $3d$ and Re $5d$ orbitals, which induces charge transfer from the Re $t_{2g}$ to Fe $t_{2g}$ orbitals.[17] Therefore, Re deficiency would reduce the amount of the charge transfer, resulting in enhancement of the unoccupied density of states (DOS) of Fe $t_{2g}$ orbitals.

In contrast to the robust trivalent Fe valence, the valence of Re ions is more significantly modified in FMI films. Figure 3b shows XAS spectra of PMM, FMM and FMI films of Re $L_3$ and $L_2$ edges. To examine the evolution of spectra more clearly, we subtracted the spectrum of an FMI film from that of a stoichiometric FMM film ($I_{FMI}$-$I_{FMM}$). The spectral weight in both $L_3$ and $L_2$ edges is shifted to higher energy in the FMI film, which indicates a higher oxidation state of Re ions.[30] It is a natural consequence of the robust trivalent $Fe^{3+}$ ions, which requires a higher oxidation state of Re ions for charge neutrality in the Fe-rich phase. Note that, the substitution of a $Re^{5+}$ ion for an $Fe^{3+}$ ion will introduce two additional $Re^{6+}$ ions to maintain the average valence of B site cations as 4+ as confirmed from the XAS data.

We found that the excess $Fe^{3+}$ ions and emergence of $Re^{6+}$ ions are keys to understand the dramatic increase of resistance of the Fe-rich phase. In this regime, there are three bonding types connecting the nearest B-site cations, i.e., $Fe^{3+}$–$Re^{5+}$, $Fe^{3+}$–$Fe^{3+}$, and $Fe^{3+}$–$Re^{6+}$. As shown in **Figure 4**a, we can draw schematics of local DOS for each bonding based on density functional theory (DFT) calculations of $A_2FeReO_6$,[31] $LaFeO_3$,[32] and $Sr_2FeWO_6$.[17] For the original $Fe^{3+}$–$Re^{5+}$ bonding in stoichiometric samples, Re $5d$ orbitals are located below Fe $3d$ orbitals. Strong hybridization and charge transfer between Fe $3d$ and Re $5d$ orbitals, therefore, induce dispersive band structures crossing the Fermi level ($E_F$), yielding the metallic behavior. On the other hand, the $Fe^{3+}$–$Fe^{3+}$ bonding found in, for example the simple perovskite $LaFeO_3$,[32] is known to result in an insulating state due to the strong Coulomb interaction. The $Fe^{3+}$–$Re^{6+}$ bonding is expected to show an insulating nature as the $Fe^{3+}$–$Re^{6+}$ bonding is the same as the $Fe^{3+}$–$W^{5+}$ bonding in the insulating $Sr_2FeWO_6$. A previous DFT calculation explained that strong hybridization between extended W $5d^1$ and O $2p$ orbitals pushes the W $5d$ orbitals above the Fe $3d$ orbitals, disturbing the charge transfer and, thereby, stabilizing an insulating ground state.[17] Similar arguments can be applied to the $Re^{6+}$ $5d^1$ orbitals, implying an insulating nature for the $Fe^{3+}$–$Re^{6+}$ bonding. Since one substitution of one $Re^{5+}$ ion with an $Fe^{3+}$ ion introduces not only one $Fe^{3+}$–$Fe^{3+}$ but also two $Fe^{3+}$–$Re^{6+}$ bondings, this modification in turn triples the effectiveness of stabilizing the insulating ground state.

Optical spectroscopy further confirms the schematic DOS shown in Figure 4a. Figure 4b shows the real part of optical conductivity $\sigma_1(\omega)$ between 1.2 and 5 eV of PMM, FMM and FMI phases measured by a spectroscopic ellipsometer. For the original $Re^{5+}$–$Fe^{3+}$ bonding of stoichiometric $Sr_2FeReO_6$, one can expect two types of optical transitions: *d-d* transitions between Re 5*d* and Fe 3*d* orbitals (labeled as *A* in Figure 4a) and *p-d* transitions from O 2*p* to Re/Fe *d* orbitals (*B* in Figure 4a). As expected, we observed two distinct spectral features, i.e., a broad spectral weight below 2.5 eV (*A*) and a strong peak near 4 eV (*B*), in $\sigma_1(\omega)$ of a FMM film (green line), consistent with previous bulk data.[33] In the FMI film, on the other hand, spectral weights of *A* and *B* were dramatically reduced while a new peak denoted as *C* emerged. The peak *C* could be assigned as transitions from O 2*p* to Fe 3*d* orbitals in the $Fe^{3+}$–$Fe^{3+}$ bonding in Figure 4a, similar to a previous optical study on $La^{3+}Fe^{3+}O_3$.[34] We did not observe any absorption expected from the $Fe^{3+}$–$Re^{6+}$ bonding, i.e. transitions from O 2*p* to Re/Fe *d* orbitals, in this accessible energy regime. As shown in the schematic diagram in Figure 4a, O 2*p* states of this bonding state is farther from $E_F$ than those of other two bonding states by the strong O 2*p* – Re $5d^1$ hybridization, which yields the corresponding optical transitions at higher energies than our instrumental limit.

Since oxides are prone to suffering from oxygen nonstoichiometry, we inspected our samples with XAS and found that the effect of oxygen vacancies is minor in stabilizing the insulating ground state. The valence of Fe (Re) ions in our cation stoichiometric films turned out to be close to 3+ (5+) in XAS spectra. We, therefore, conclude that the oxygen background pressure as low as 20 mTorr is sufficiently high enough to ensure the full oxidation.

In summary, we discovered a new room-$T$ ferromagnetic insulating state by deliberately controlling the ratio of two B-site cations in the double perovskite $Sr_2FeReO_6$. A single growth parameter $P_{O2}$ was used to control the cation ordering and ratio. Particularly in the ordered/Fe-rich growth regime, we found intriguing FMI states which were attributed to excess $Fe^{3+}$ and emergent $Re^{6+}$ ions. Our results provide a new promising route to developing oxide ferromagnetic insulators, and further studies on spin filter, magnetic proximity, and quantum anomalous Hall effects with this new FMI may enable the development of energy efficient quantum electronic and spintronic devices.

**Experimental Section**

*Synthesis of $Sr_2FeReO_6$ films*: We used pulsed laser deposition to grow high-quality $Sr_2FeReO_6$ films in a wide range of growth parameters, such as temperature ($T$ = 650 – 775 ºC), oxygen partial pressure ($P_{O2}$ = 0.1 – 100 mTorr), laser fluence (1–2 J/cm$^2$), and laser repletion rate (1–10 Hz). Among the growth parameters, oxygen partial pressure turned out to be the most critical for cation ordering and stoichiometry control. Therefore, the samples presented here were grown at fixed conditions after extensive optimization processes – $T$ (775 ºC), laser fluence (1.5 J/cm$^2$), and frequency (5 Hz) – except the oxygen pressure. All the films used in this work are 30-40 nm in thickness.

*Basic characterization*: High-resolution four circle x-ray diffraction, superconducting quantum interference device, and physical property measurement system were used to check the crystallographic, magnetic, and transport properties of thin films, respectively.

*Scanning transmission electron microscopy*: Cross-sectional specimens for scanning transmission electron microscopy imaging were prepared using ion milling after mechanical thinning and precision polishing. High-angle annular dark-field imaging was carried out in a Nion UltraSTEM100 operated at 100 keV. The microscope is equipped with a cold field-emission gun and an aberration corrector for sub-angstrom resolution. An inner detector angle of 78 mrad was used for imaging. The convergence semi-angle for the electron probe was set to 30 mrad.

*Rutherford back-scattering spectrometry*: Rutherford back-scattering spectrometry technique was used to determine the Re/Fe ratio in the films. A parallel 2.0 MeV He beam was employed in these measurements with a detector located underneath the He beam at a backscattering angle of 168° and solid angle of $2.3 \times 10^{-3}$ sr to have a combining advantage of optimized mass resolution and depth resolution.[35]

*X-ray absorption spectroscopy*: Fe $L$–edge spectra were obtained at the beamline 4–ID–C of Advanced Photon Source at 50 K with total electron and fluorescence yield mode. The Re $L$–edge spectra were taken at the beamline 4-ID-D of Advanced Photon Source with an energy dispersive detector by collecting the Re $L_\alpha$ and Re $L_\beta$ fluorescence lines, respectively for the $L_3$ and $L_2$ edges.

*Spectroscopic ellipsometry*: By using a M-2000 ellipsometer (J.A. Woollam Co.), two ellipsometric parameters, $\psi$ and $\Delta$, for thin films and a bare $SrTiO_3$ substrate were obtained at room temperature. A two-layer model composed of the substrate and film was constructed to determine optical constants of the films.

**Acknowledgements**

This work was supported by the U.S. Department of Energy (DOE), Office of Science, Basic Energy Sciences, Materials Sciences and Engineering Division. This research used resources of the Center for Nanophase Materials Sciences (spectroscopic ellipsometry), which are DOE Office of Science User Facilities. Use of the Advanced Photon Source was supported by the US DOE, Office of Science, under Contract No. DE-AC02-06CH11357. L.N. was supported for the RBS work by the University of Tennessee Governor's Chair program.

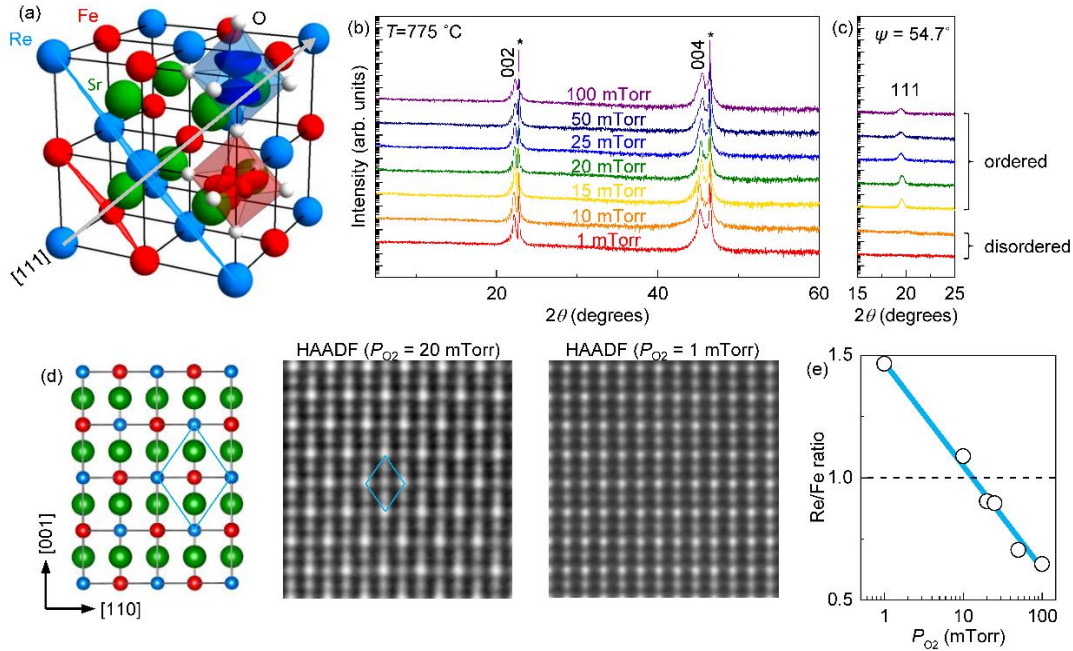

**Figure 1.** Epitaxial growth of highly cation-ordered epitaxial $Sr_2FeReO_6$ films. a) Schematic of the double perovskite $Sr_2FeReO_6$. Blue and red planes are alternating atomic planes of Re and Fe along the [111] direction, respectively. The [111] direction is indicated with a gray line. Spin-orbit coupled 5$d$ wavefunction and localized 3$d$ wavefunction are visualized inside octahedra. b) XRD $\theta$–$2\theta$ scans for $Sr_2FeReO_6$ films on $SrTiO_3$ grown at 775 °C under different $P_{O2}$. Asterisks indicate peaks from the $SrTiO_3$ substrate. c) Off-axis XRD $\theta$–$2\theta$ scans near the $Sr_2FeReO_6$ 111 peak. Its appearance confirms the Fe/Re ordering. d) A schematic of the projected double perovskite structure along the [1-10] direction (left), a HAADF image of a cation-ordered $Sr_2FeReO_6$ film grown at 20 mTorr (middle), and a HAADF image of a disordered $Sr_2FeReO_6$ film grown at 1 mTorr (right). The blue diamond in the schematic and HAADF image highlights the ordered Re ions. e) Re/Fe ratio obtained from RBS measurements as a function of $P_{O2}$. The blue line is a guide for eyes.

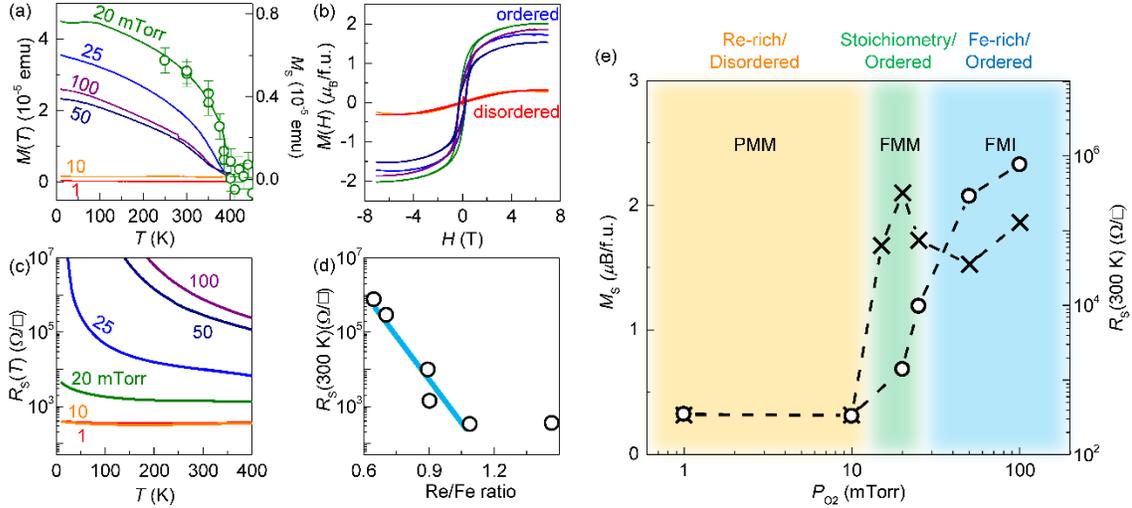

**Figure 2.** Metal-insulator transition with robust ferromagnetism above room temperature. a) $M(T)$ curves with field cooling and measuring at 1000 Oe. Green open circles are $M_S(T)$ data obtained by a high-$T$ oven setup for an ordered/stoichiometric $Sr_2FeReO_6$ film ($P_{O2}$ = 20 mTorr) to determine the $T_c$. Due to the instrumental size limitation of the sample holder for the high-$T$ oven setup, isothermal magnetization curves were obtained between 4 and –4 kOe. The data above 1 kOe and below –1 kOe were fit with a line to estimate the $M_S$ of the sample at $T$s between 250 and 500 K. b) $M(H)$ curves at 10 K. The magnetic ordering in $Sr_2FeReO_6$ is strongly suppressed in cation-disordered samples. c) Semi-log plot of $R_S(T)$: A clear metal-insulator transition was observed as the $P_{O2}$ increases. d) $R_S(T)$ versus Re/Fe ratio, which shows a strong correlation between Re deficiency and electronic ground state. e) Phase diagram of $Sr_2FeReO_6$ films with different $P_{O2}$. The cross symbols indicate the $M_S$, while open circles are $R_S$(300 K) as a function of $P_{O2}$. PMM, FMM, and FMI ground states are stabilized in disordered/Re-rich, ordered/stoichiometric, and ordered/Fe-rich phases, respectively.

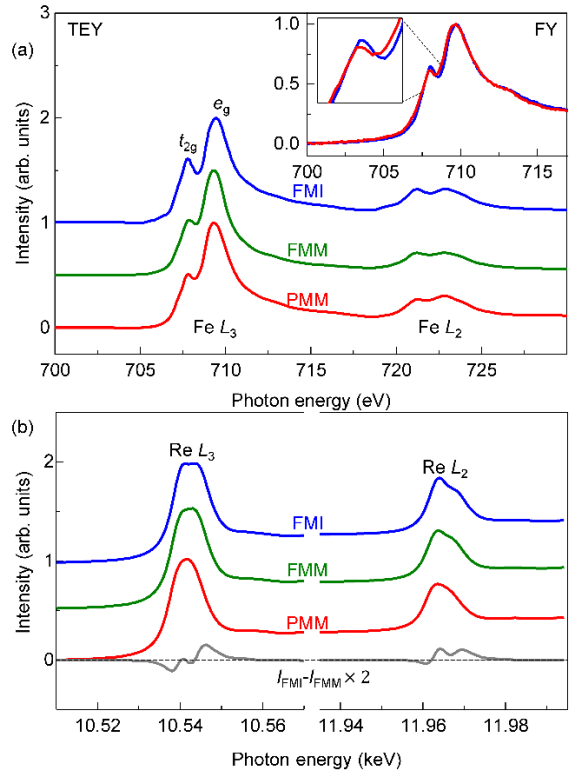

**Figure 3.** XAS spectra. a) Fe *L*- and b) Re *L*- edge XAS spectra of PMM, FMM and FMI films. Fe *L*–edge spectra show a small change among films except a slight increase of the peak intensity of unoccupied $t_{2g}$ orbitals in the FMI film. This result indicates the reduced hybridization between Re 5*d* and Fe 3*d* orbitals. The inset of (a) shows FY spectra near the Fe $L_3$ edge of PMM and FMI films, validating that the observed spectral change in TEY is not coming from the surfaces. In Re *L*–edge spectra, on the other hand, the spectral weight moves to a higher energy in the FMI film due to the emergence of $Re^{6+}$.

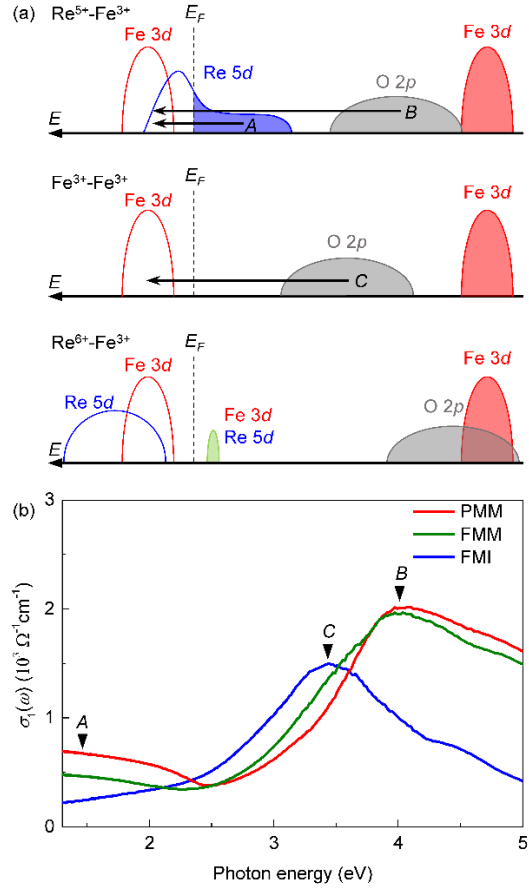

**Figure 4.** Electronic structures and optical conductivity of $Sr_2FeReO_6$ films. a) Local DOS of three types of bonding in an Fe-rich $Sr_2FeReO_6$ film based on reported DFT calculations.[17, 31, 32] The expected optical transitions in $Fe^{3+}$–$Re^{5+}$ and $Fe^{3+}$–$Fe^{3+}$ bonding were denoted as *A*, *B* and *C*. b) $\sigma_1(\omega)$ of PMM, FMM and FMI films: The three expected optical transitions are observed in the spectra of FMM and FMI films.